\documentclass[aps,pre,preprint,showpacs,superscriptaddress,longbibliography]{revtex4-2} 
\usepackage[caption=false]{subfig}
\usepackage{graphicx}
\usepackage{dcolumn}
\usepackage{bm}
\usepackage{float}

\makeatletter
\let\newfloat\newfloat@ltx
\makeatother

\usepackage{algpseudocode}
\usepackage{algorithm}

\usepackage{float}
\usepackage[english]{babel}
\usepackage{tabularx}
\usepackage{hhline}
\usepackage{amsmath}
\usepackage[colorlinks=true, allcolors=blue]{hyperref}
\usepackage{tikz}
\usepackage{amssymb}
\usepackage{xcolor}
\usepackage{color}

\usetikzlibrary{shapes.geometric, arrows}
\tikzstyle{startstop} = [rectangle, rounded corners, minimum width=3cm, minimum height=1cm,text centered, draw=black, fill=red!30]
\tikzstyle{io} = [trapezium, trapezium left angle=70, trapezium right angle=110, minimum width=3cm, minimum height=1cm, text centered, draw=black, fill=blue!30]
\tikzstyle{process} = [rectangle, minimum width=3cm, minimum height=1cm, text centered, draw=black, fill=orange!30]
\tikzstyle{decision} = [diamond, minimum width=3cm, minimum height=1cm, text centered, draw=black, fill=green!30]
\tikzstyle{arrow} = [thick,->,>=stealth]

\begin{document}

\title{A Framework Based on Symbolic Regression Coupled with eXtended Physics-Informed Neural Networks for Gray-Box Learning of Equations of Motion from Data}

\author{Elham Kiyani}
\affiliation{Department of Mathematics, The University of Western Ontario, 1151 Richmond Street, London, Ontario, N6A~5B7, Canada}
\affiliation{The Centre for Advanced Materials and Biomaterials (CAMBR), The University of Western Ontario, 1151 Richmond Street, London, Ontario, N6A~5B7, Canada} 
\author{Khemraj Shukla}
\affiliation{Division of Applied Mathematics, Brown University, 182 George Street, Providence, RI 02912, USA}
\author{George Em Karniadakis}
\affiliation{Division of Applied Mathematics, Brown University, 182 George Street, Providence, RI 02912, USA}
\author{Mikko Karttunen}
\affiliation{The Centre for Advanced Materials and Biomaterials (CAMBR), The University of Western Ontario, 1151 Richmond Street, London, Ontario, N6A~5B7, Canada} 
\affiliation{Department of Physics and Astronomy,
 The University of Western Ontario, 1151 Richmond Street, London,
 Ontario,  N6A\,3K7, Canada}
\affiliation{Department of Chemistry, The University of Western Ontario, 1151 Richmond Street, London, Ontario, N6A~5B7, Canada}

\date{\today}
\begin{abstract}
We propose a framework and an algorithm to uncover the unknown parts of nonlinear equations directly from data. The framework is based on eXtended Physics-Informed Neural Networks (X-PINNs), domain decomposition in space-time, but we augment the original X-PINN method by imposing flux continuity across the domain interfaces. The well-known Allen-Cahn equation is used to demonstrate the approach. The Frobenius matrix norm is used to evaluate the accuracy of the X-PINN predictions and the results show excellent performance. In addition, symbolic regression is employed to determine the closed form of the unknown part of the equation from the data, and the results confirm the accuracy of the X-PINNs based approach. To test the framework in a situation resembling real-world data, random noise is added to the datasets to mimic scenarios such as the presence of thermal noise or instrument errors. The results show that the framework is stable against significant amount of noise. As the final part, we determine the minimal amount of data required for training the neural network. The framework is able to predict the correct form and coefficients of the underlying dynamical equation when at least 50\% data is used for training.  

\end{abstract}
\maketitle

\section{Introduction}

Partial differential equations (PDEs) are  commonly used for modeling the evolution of dynamical systems in, e.g., fluid dynamics, heat transfer, financial derivatives, chemical reactions and phase tranformations. From the physical perspective, one of the main problems is constructing models that contain all the relevant information about the system at hand. This involves, for example, identifying the order parameters, relevant symmetries and possible couplings, and their nature, between the order parameters.
Machine learning (ML) offers a method for automated construction of models directly from experimental or other data: it can be used to identify PDEs from data without prior/or only with partial knowledge about the underlying physics.

Being able to predict the PDEs from data involves training a neural network to recognize patterns in the data set(s), and then using the learned network to identify the underlying PDEs. Various methods have been proposed including PDE-Net~\cite{long2018pde,long2019pde}, neural networks~\cite{kiyani2022machine,qu2022learning}, Gaussian processes~\cite{chen2021solving,lee2020coarse}, 
and the sparse identification of nonlinear dynamics (SINDy) algorithm~\cite{kaiser2018sparse, fukami2021sparse, hoffmann2019reactive,fasel2021sindy, shea2021sindy}. However, when dealing with sparse and high-dimensional data, these methods may not be sufficient to obtain a high level of accuracy. To circumvent these issues, Raissi et al.~\cite{raissi2019physics} proposed Physics-Informed Neural Networks (PINNs), which utilize a novel approach that incorporates physics-based constraints into the loss function allowing for accurate predictions in complex systems with varying initial and boundary conditions. PINNs are capable of discovering unknown equations and solving them~\cite{karniadakis2021physics}. 

Their leading idea is to integrate the fundamental physical principles of a system with neural networks. They are also flexible in the sense that they can handle complex and non-convex geometries, and different boundary and initial conditions. The loss function for a PINN incorporates data fitting, residuals of the PDEs (computed using automatic differentiation), initial and boundary conditions. The network parameters are updated during training to minimize the loss function, resulting in a solution that meets the constraints applied in the loss function. 

One of the key benefits of using PINNs, as compared to other ML techniques, is their ability to effectively learn from limited data, while also incorporating prior knowledge of the system being studied. This prior knowledge is incorporated into the loss function of the network, and it helps to guide the network towards solutions that are physically plausible and consistent with the known properties of the system being studied.  This allows the network to make more accurate predictions even with limited training data. PINNs possess the ability to tackle both forward and inverse problems, and identifying unknown parameters in  differential equations based on observed data. To achieve this, the available data is introduced into the network's training process as solutions to the PDEs, allowing it to learn and approximate the underlying physics of the system, and subsequently recognize the unknown parameters. This feature is beneficial in instances where the governing equations are only partially known, and the aim is to deduce the parameters based on gathered experimental or observational data~\cite{raissi2019physics, lou2021physics, shukla2020physics}.

Since the introduction of PINNs, various extensions such as Physics-Informed Attention-Based Neural Networks, (PIANNs)~\cite{rodriguez2021physics}, Generative Adversarial Physics-Informed Neural Networks (GA-PINNs)~\cite{li2022revisiting}, Graph Convolutional Networks (GCNs)~\cite{gao2022physics}, and Bayesian Physics-Informed Neural Networks (B-PINNs)~\cite{yang2021b} have been developed to enhance performance and to extend the applicability of PINNs to different problems. In this work, we focus on the eXtended Physics-Informed Neural Networks (X-PINNs)~\cite{jagtap2021extended,SHUKLA2021110683}. They involve generalized space-time domain decomposition in order to provide computationally efficient solutions to PDEs across large spatial and temporal domains. In this approach, the domain is first split into smaller subdomains. Then, the PDEs are solved in each subdomain using PINNs and at the interfaces wicertain continuity conditions are imposed as soft-constraints in the loss function. This allows X-PINNs to use large neural networks without the common problem of overfitting. X-PINNs also reduce the computational cost associated with training due to their implicit concurrent implementation.

In this study, we propose a framework combining X-PINNs with data-driven methods to uncover the non-linear term of the underlying PDE, while assuming the presence of a Laplacian term as a diffusion operator. The well-known Allen-Cahn model is used as the test case~\cite{allen1972ground}. Our study presents a promising approach for gray-box learning of equations with X-PINNs. Gray-box learning refers to discovering equations when only partial knowledge of the equation is available and it combines the strengths of both white-box and black-box learning. It can be particularly useful in situations where the known parts of the equation provide valuable insights into the behavior of the system, despite the complete equation being unknown~\cite{kemeth2023black, yaghoubi2019gray}. For instance, in the current study, the Laplacian term represents the known part of the equation, while the non-linear term is unknown. After discovering the non-linear term of the gray-box Allen-Cahn equation using X-PINN, we feed the discovered term and the data from phase-field simulations into a symbolic regression model to predict the explicit mathematical formula of the unknown term. Symbolic regression is an ML technique used to discover the explicit mathematical expressions or equations that best fit a given dataset~\cite{billard2002symbolic}. The framework is implemented using Python, and utilizes the Tensorflow (version 2.0) deep learning framework for its efficient automatic differentiation capabilities~\cite{abadi2016tensorflow}.

The rest of this article is structured as follows: Section~\ref{Phase-field model} provides an overview of the phase-field approach and data preparation. Section~\ref{X-PINN} gives a brief summary of the PINNs and X-PINNs, followed by a presentation of X-PINN results and a comparison of the predictions using PINNs and X-PINNs. Section~\ref{Symbolic regression} presents the symbolic regression results. The performance of the framework for noisy data is discussed in Section~\ref{Noisy data}. In Section~\ref{Reducing training data sets}, we examine the framework's performance for different sizes of training data sets to investigate the amount of optimal data required for training. Finally, a summary of the current work is provided in Section~\ref{Conclusion}.

%

\section{Phase-Field Modeling}\label{Phase-field model}

In phase-field modeling the time evolution of the order parameter ($U(\vec{x},t)$) is described by a time-dependent PDE~\cite{Provatas2010-hc}. The order parameter takes the value zero in the disordered, or high-temperature phase, and a small finite value in the ordered, or low-temperature phase. The order parameter field is called the ``phase-field" and the order parameter itself can be a scalar, vector or even a tensor depending on the nature of the system at hand~\cite{Hohenberg1977-el,Provatas2010-hc, colli2012free}. The typical way of constructing such models is by postulating a phenomenological Ginzburg–Landau free energy, or Lyapunov functional, in terms of a gradient expansion of the order parameter and taking a functional derivative~\cite{Hohenberg1977-el,Provatas2010-hc}. The terms that are included from the expansion must obey the symmetries of the system. In cases when a free energy functional cannot be constructed, one typically postulates an equation of motion. This is, for example, the case with reaction-diffusion models~\cite{Pearson1993-pj,Leppanen2004-lb}. Systems may also have several order parameters that are coupled, examples include such diverse systems as magnetocrystallinity~\cite{Faghihi2013-tk} and cell migration~\cite{Najem2016-aq}. Recently, open source software for phase-field simulations has also started to emerge, see e.g. Refs.~\cite{Hong2020-fc, Silber2022-pz}.

\begin{figure}[!tb]
    \centering
    \includegraphics[width=\textwidth]{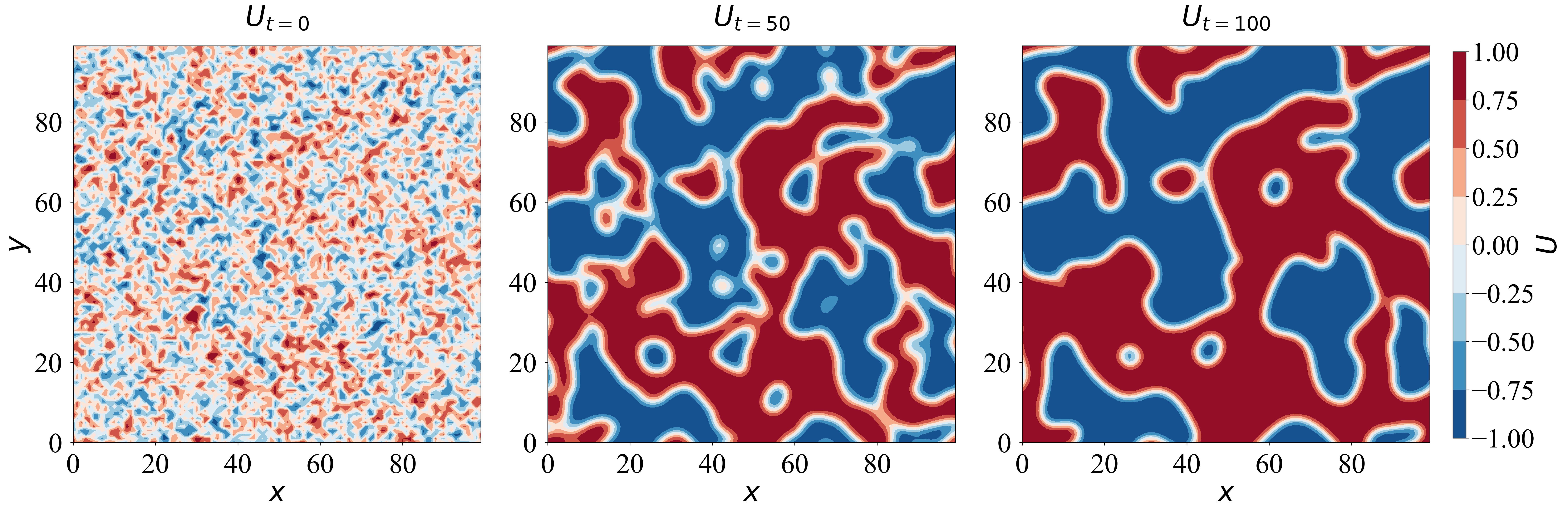}
    \caption{Snapshots
    from a simulation of the Allen--Cahn model, Equation~\eqref{eq:modela}, at $t=0$, $t=50$, and $t=100$. The simulation was performed using dimensionless units, and on a uniformly discretized grid of size $n_x \times n_y = 100 \times 100$ with a spatial resolution of $\Delta x = \Delta y =1.0$. A time step of $\Delta t = 0.1$ was used. Periodic boundary conditions were applied and the initial configuration was randomly generated from a uniform distribution. 
    }
    \label{fig:Allen--Cahn}
\end{figure}

We utilize the well-known Allen--Cahn model~\cite{allen1972ground,Chen2002-ej}. This model was first introduced by Stuart Allen and John Cahn in 1972 to describe solidification dynamics in binary alloys by employing a non-conservative scalar order parameter~\cite{allen1972ground}. Its wide-ranging applications in solidification include, for example, dendritic growth and pattern formation. For broader discussions see, e.g. Refs.~\cite{Chen2002-ej,li2012phase, nepomnyashchy2015coarsening}. The model describes the evolution of a phase interface via a PDE that accounts for the thermodynamic driving force for phase separation, as well as kinetic effects arising from diffusion and surface tension. The equation of motion for the order parameter $U \equiv U(\vec{x},t)$ can be given in a dimensionless form as~\cite{Provatas2010-hc}
\begin{equation}
\dfrac{\partial U}{\partial t} =  -M \left(\nabla^{2} U + a_{2} U - a_{4 }U^3\right)\,, 
\label{eq:modela}
\end{equation}
where the rate of change of the order parameter over time is determined by the mobility coefficient $M$. It is related to the interfacial energy, and it controls the speed of interface propagation and the wavelength of resulting patterns. Here we set $M = 1$. The constants $a_{2}$ and $a_{4}$ determine the shape and behavior of the free energy density of the system. We set $a_{2} = a_{4} = 1$ in our simulations. 
Figure~\ref{fig:Allen--Cahn} displays snapshots at three different time steps ($t=0$, $t=50$, and $t=100$) to illustrate the dynamics of the Allen--Cahn model.

In addition to traditional numerical methods, recent developments in machine learning have shown promise in solving the Allen-Cahn equation as a forward problem. The PINN framework has been demonstrated to be highly accurate and efficient for solving the equation~\cite{raissi2019physics,mattey2022novel,wight2020solving}. In addition, the Laplacian part has elliptic regularity, resulting in fast convergence of PINNs~\cite{shin2020convergence}.  Another emerging field is physics-informed machine learning (PIML), which combines traditional numerical methods with ML to solve physical problems such as the Allen-Cahn equation.
In other recent studies, data driven deep learning frameworks have been proposed and shown to significantly accelerate the traditional numerical methods for solving phase-field equations~\cite{li2023phase, oommen2022learning, kiyani2022machine}. 

Here, numerical simulations were conducted using a two-dimensional grid with dimensions $[n_x \times n_y] = [100 \times 100]$ and grid spacing of $[\Delta x, \Delta y] =[1, 1]$. The simulations were run from $t=0$ to $t=10$ with a time step of $\Delta t = 0.1$, resulting in a total of $n_t=100$ time steps. Periodic boundary conditions were applied and uniform random distribution was used for initial conditions. Figure~\ref{fig:Allen--Cahn} shows snapshots from a simulation starting from random initial conditions.

\section{Extended Physics-Informed Neural Network (X-PINN)}\label{X-PINN}

PINNs are a type of ML algorithm that can accurately solve differential equations by incorporating the known physics (e.g., PDEs, ODEs, integro-differential equations) of the problem into a neural network architecture as soft constraints~\cite{raissi2019physics,karniadakis2021physics,cai2021physics,stiasny2021physics}. The network is trained to minimize the loss functions constructed by computing the misfit between the data, initial and boundary conditions, and residuals of the PDE. To minimize the loss function, a first order optimizer (ADAM) or a combination of first and second order optimizer (ADAM + L-BFGS) is typically used~\cite{fatima2020enhancing,mustapha2021comparative}; L-BFGS stands for the limited-memory Broyden–Fletcher–Goldfarb–Shannon algorithm. This approach allows PINNs to capture the underlying physics in the data and to make predictions. PINNs are particularly suited for modeling non-linear relationships, and handling multiple physical processes~\cite{stiasny2021physics, rodriguez2021physics, ma2022preliminary, zhu2021machine, choi2022physics, my2022physics, broeckhoven2017has, bastek2023physics, cai2021physics, laubscher2021simulation}.

In general, using a single PINN may not be sufficient to accurately capture chaotic, highly non-linear, and complex solutions. Increasing the depth of the neural network may seem like an obvious solution, but it can lead to several issues. Firstly, using deeper networks with sparse data can result in over-parameterization, which can lead to over-fitting. This means that the network may fit the training data too closely and fail to generalize to new data. Secondly, deep neural networks can result in a complex and highly non-convex loss landscape, which can make it difficult to optimize the network. This can lead to issues such as getting stuck in local minima and slow convergence. Lastly, using deeper networks can be computationally intractable, which can limit their usefulness in practical applications. To circumvent these issues, Jagtap et al.~\cite{jagtap2021extended} proposed X-PINNS which are primarily based on the domain decomposition approach frequently adopted in the classical numerical methods.

The X-PINN approach extends PINNs by offering space-time domain decomposition, which can be useful for problems with irregular, and non-convex geometries. X-PINNs aim to improve the accuracy and efficiency of solving and discovering PDEs by introducing additional physics-inspired constraints. In this approach, the domain is decomposed into smaller subdomains both in space and time, and separate neural networks are trained for each subdomain. The solutions at the spatio-temporal interfaces of the subdomains are unified by enforcing residual continuity. This allows for more efficient and accurate modeling of the physics in each subdomain, leading to an improved overall accuracy of the solutions.

As discussed in Section~\ref{Phase-field model}, we use the Allen-Cahn model as the test case. The underlying assumption is that the Laplacian term of Equation~\eqref{eq:modela} is considered to be the known part of the equation and used as prior knowledge during the training of X-PINNs. Thus, the objective is to determine the function $F(U)$ such that
\begin{equation}
\dfrac{\partial U}{\partial t} =  \nabla^{2} U + F(U)\,.
\label{eq:F}
\end{equation}

\begin{figure}[!tbh]
    \centering
    \includegraphics[width=1.05\textwidth]{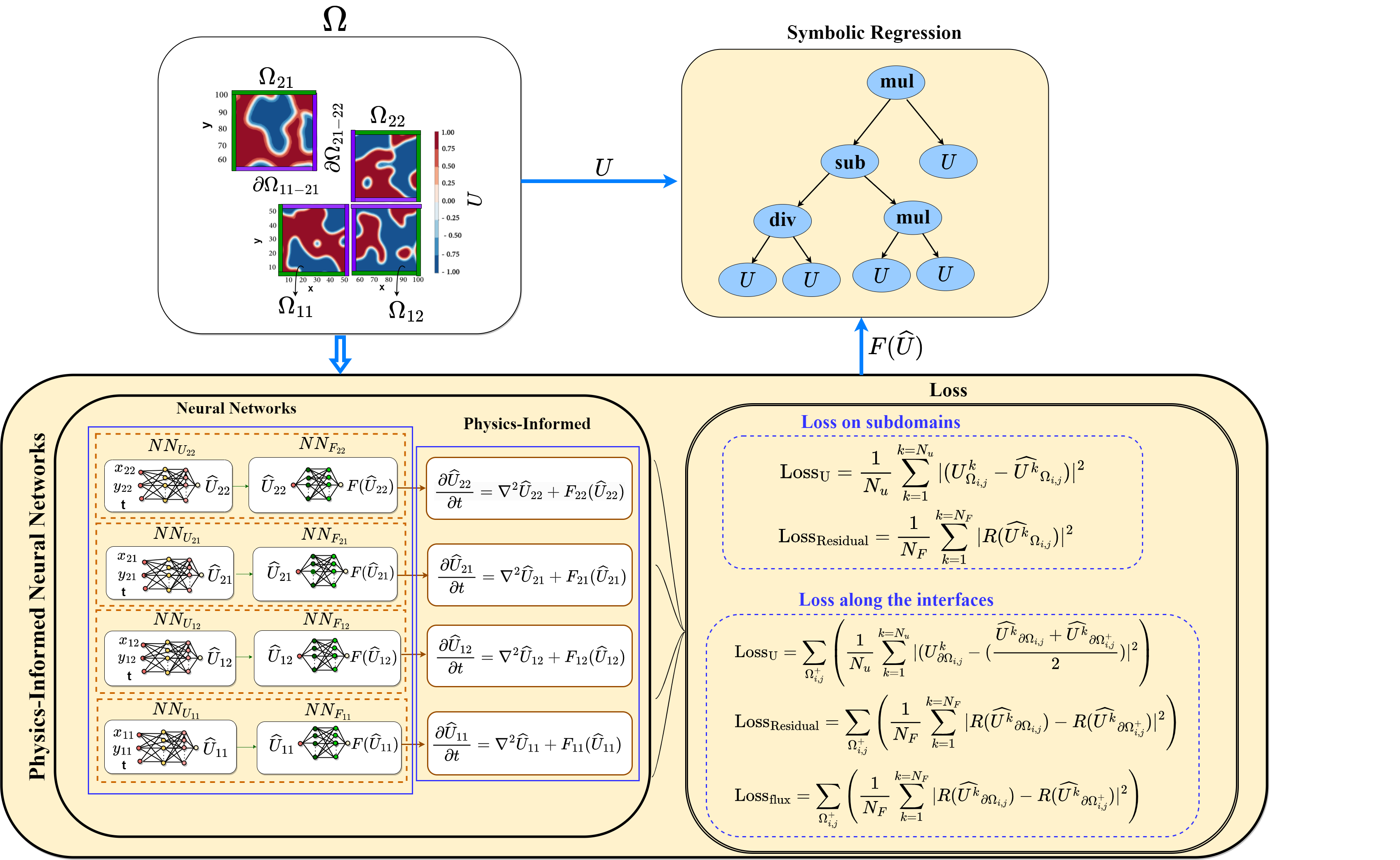}
    \caption{The X-PINN methodology for discovering the Allen--Cahn model with four subdomains, ${\Omega_{11}}$ ($0 \leq x , y \leq 50$), ${\Omega_{12}}$ ($50 \leq x \leq 100$ and $0 \leq y \leq 50$), ${\Omega_{21}}$ ($0 \leq x \leq 50$ and $50 \leq y \leq 100$), and ${\Omega_{22}}$ ($50 \leq x , y \leq 100$) involves several steps. Four sub-PINNs corresponding to the four subdomains are composed, each consisting of two sub-networks, $NN_{U}$ and $NN_{F}$, and a physics-informed part. $NN_{U}$ takes inputs $x$, $y$, and $t$ at each subdomain to predict the output $\widehat{U}$. The output $\widehat{U}$ is then fed into a second network $NN_{F}$ to predict the output $F(\widehat{U})$. Using the predicted $\widehat{U}$ and $F(\widehat{U})$, the physics-informed part creates Equation~\eqref{eq:F}. The loss function is composed of two categories: 1) loss on subdomains and 2) loss along the interfaces, where $\mathrm{Loss_{U}}$ and $\mathrm{Loss_{residual}}$ minimize data mismatch and residual on each subdomain, respectively. Additionally, the average solution continuity term and the residuals across the subdomain interfaces are included in the loss function, along with $\mathrm{Loss_{flux}}$, which represents the normal flux continuity term. After minimizing the loss function, the next step involves feeding $U$ and predicted $F(\widehat{U})$ into symbolic regression to predict the general form of $F$ as a function of $U$.}
    \label{fig:schematic_XPINN}
\end{figure}

Figure~\ref{fig:schematic_XPINN} shows a schematic diagram of the proposed framework to discover the  unknown function $F(U)$ in Equation~\eqref{eq:F}. To train the X-PINN, the domain $\Omega$ is divided into four subdomains such that, $\Omega_{11}$ with $0 \leq x \leq 50$ and $0 \leq y \leq 50$, $\Omega_{12}$ with $50 \leq x \leq 100$ and $0 \leq y \leq 50$, $\Omega_{21}$ with $0 \leq x \leq 50$ and $50 \leq y \leq 100$, and $\Omega_{22}$ with $50 \leq x \leq 100$ and $50 \leq y \leq 100$.

In this study, X-PINNs are used with four sub-PINNs, each consisting of two sub-networks. The sub-networks $NN_{U_{i,j}}$, for $i,j$ in ${1,2}$, predict $U_{i,j}$ as a function of $x$, $y$, and $t$, and the sub-networks $NN_{F_{i,j}}$ are used to predict the corresponding $F_{i,j}(U)$. The network architectures for both sub-networks are shown in Table~\ref{tab:networks}. The architectures of the networks $NN_{U_{i,j}}$ include six dense layers, each having $20$ neurons, while the networks $NN_{F_{i,j}}$ comprise of four dense layers with $20$ neurons in each layer. The networks were trained using a learning rate of $10^{-3}$ for $300,000$ epochs. Each subdomain has $n_t = 100$ snapshots randomly split into training and test sets with an $80:20$ ratio.

\begin{table}
\centering
\begin{tabular}{c|c|c|c|c}
\hline
Networks  & \# of layers & Layer type  & Neurons in each layer &  Activation function \\ 
\hline
$NN_{U_{i,j}}$   & 6 & dense and fully connected  & $20$  & $\tanh$ \\ 
\hline
$NN_{F_{i,j}}$   & $4$ & dense and fully connected & $20$  & $\tanh$ \\ 
\hline
\end{tabular}
\label{tab:networks}
\caption{The neural network architectures in both sub-networks. Networks $N_{U_{i,j}}$ consist of $6$ layers with $20$ neurons in each layer.
The $N_{F_{i,j}}$ networks are comprised of $4$ layers and $20$ neurons in each layer. The networks were trained with learning rate of $10^{-3}$ for $300,000$ epochs. Each subdomain has $n_t = 100$ snapshots wrandomly split into training and testing with a $80:20$ ratio.}
\end{table}

As shown in Figure~\ref{fig:schematic_XPINN}, the first set of networks $N_{U_{i,j}}$ takes $x_{i,j}$, $y_{i,j}$, and $t$ as inputs and predicts $\widehat{U}_{i,j}$ as a function of these inputs. The predicted $\widehat{U}_{i,j}$ is then fed through the second set of networks $N_{F_{i,j}}$ to discover the function $F(\widehat{U}_{i,j})$. Additionally, there are four interfaces $\partial \Omega_{11-12}$, $\partial \Omega_{11-21}$, $\partial \Omega_{12-22}$, and $\partial \Omega_{21-22}$, between two neighboring subdomains. $\partial \Omega_{11-12}$ and $\partial \Omega_{11-12}$ are depicted in Figure~\ref{fig:schematic_XPINN} as examples which $\partial \Omega_{11-12}$ is a common boundary between subdomains $\Omega_{11}$ and $\Omega_{12}$, and has $x=50$ and $0 \leq y \leq 50$. Therefore, each subdomain has two interfaces and two boundaries, marked by purple and green, respectively, in Figure~\ref{fig:schematic_XPINN}.

\begin{algorithm}[!htb]
\caption{Pseudo-algorithm for gray-box learning of the Allen-Cahn equation}
\label{alg:GBAC}
    \begin{algorithmic}
        \Require $U$, $x$, $y$, $t$: sampled using Latin hypercube sampling     
        \Require $N_d$: Number of sub-domains to partition $\Omega$ in subdomains 
        \Require $i_d$: iterator for subdomains
        \Require $\Theta_i=\{W_i, b_i\}_{i=1}^{N_d}$ \Comment{Initialize trainable parameters}
        \Require $\epsilon$: $10^{-5}$ \Comment{Initialize training convergence criteria}
        \Require $N$: Total number of iterations \Comment{Number of iterations}
        \Require Update $\in$ \{ Adam, L-BFGS\} \Comment{Set of optimizers} 
        \Require $\mathcal{F}_{{\widehat U}}(\bm{\Theta_i}), \mathcal{F}_{F_ {\widehat U}}(\bm{\Theta_i})$ \Comment{Initialize neural networks for each subdomain for $U$ and $F(U)$}
        \Require $n,~i_d,~\mathcal{L}$: iteration counters and initial loss  \Comment{Initialize}
        \ForAll{{$\mathit{i_d}\in N_d$}} \Comment{Loop 1: X-PINN Training}
        \While{{$\mathcal{L} > \epsilon$ and $n < N$}} 
           \State $U_{i_d},~F(\widehat U)_{i_d} \gets \mathcal{F}_{{\widehat U} }(\bm{\Theta_{i_d}}),~ \mathcal{F}_{F_ {\widehat U}}(\bm{\Theta_{i_d}})$
           \State $\mathcal{L}_{i_d} \gets \mathcal{L}_{U_{i_d}} + \mathcal{L}_{F({\widehat U})_{i_d}} + \mathcal{R}(U)_{i_d} $ \Comment{$\mathcal{R}$ is residual loss} 
           \State $\bm{\Theta}_{i_d} \gets \text{Update}({\bm{\Theta}}) $
           \State $n \gets n + 1$
          \EndWhile
         \State $i_d \gets i_d + 1$
        \EndFor
         \State $U_{i_d}, F(\widehat U)_{i_d} \gets \text{Loop: 1}$
         \Require $F(\widehat U)_{i_d}: \alpha_{i_d}U^m_{i_d} + \beta_{i_d}U^n_{i_d}$
         \Require $\epsilon_{rR_{Tol}}$
         \Require $\epsilon_{Sr} = 100$
         \Comment{Tolerance for symbolic regression}
        \ForAll{{$\mathit{i_d} \in N_d$}} \Comment{Loop 2: Symbolic regression}
        \While{$\epsilon_{Sr}$ $<$ $\epsilon_{Sr_{Tol}}$}
           \State $ \{\alpha_{i_d},\beta_{i_d}, m_{i_d}, n_{i_d}, \epsilon_{Sr}\} \gets \text{Symbolic Regressor}((U_{i_d}, F(\widehat U)_{i_d}))$
           \EndWhile
        \EndFor        
    \end{algorithmic}
\end{algorithm}

It is important to note that in X-PINNs, the loss functions are defined separately for each subdomain. Each subdomain has the same terms as the standard PINN loss function, that is, a data-fitting term and the residuals of the PDE expressed in Equation~\eqref{eq:F}. Let $N_{u}$ and $N_{F}$ denote the number of training and residual data points, respectively. The unknown function $F$ can be obtained by minimizing the mean squared error loss
\begin{equation}\label{eq:MSE-data}
\mathrm MSE_{U_{\Omega_{i,j}}} = \frac{1}{N_{u}} \sum_{k=1}^{k=N_{u}} | (U^k_{\Omega_{i,j}} - \widehat{U^k}_{\Omega_{i,j}}) | ^{2},
\end{equation}
\begin{equation}\label{eq:MSE-equation}
\mathrm MSE_{\mathrm{Residual}_{\Omega_{i,j}}} = \frac{1}{N_{F}} \sum_{k=1}^{k=N_{F}} | R(\widehat{U^k}_{\Omega_{i,j}}) | ^{2},
\end{equation}
where $\mathrm MSE_{U_{\Omega_{i,j}}}$ is a data-fitting term, and $U_{\Omega_{i,j}}$ and $\widehat U_{\Omega_{i,j}}$ are the simulation data and the predicted solution by sub-PINNs over subdomain $\Omega_{i,j}$. $\mathrm MSE_{Residual_{\Omega_{i,j}}}$ is the loss for the PDE residual that enforces the governing Equation~\eqref{eq:F} over subdomains. For the predicted $\widehat U_{\Omega_{i,j}}$ it is defined as
\begin{equation}
R(\widehat {U}_{\Omega_{i,j}}) = \dfrac{\partial \widehat {U}_{\Omega_{i,j}}} {\partial t} -  \nabla^{2} \widehat {U}_{\Omega_{i,j}} - F(\widehat {U}_{\Omega_{i,j}})\,.
\label{eq:R}
\end{equation}
%

Subsequently, each subdomain's loss function is modified to enforce flux and residual continuity at the interface. This binds the subdomains together and ensures unique solutions at the interface between two neighboring subdomains. It is to be noted that in the original implementation of X-PINN \cite{jagtap2021extended, shukla2020physics}, only continuity of the solutions and residuals are enforced at the interfaces of two neighboring subdomains. In this study, we augment the loss function with flux continuity along with continuity of the residuals and the solution. This enables the model to consider physical constraints that are not explicitly included in the training data. By doing so, the model's accuracy and the rate of convergence are improved, enabling it to determine more precise predictions regarding the interface behavior. The loss terms representing the continuity of solutions, residual and flux at the interface of two subdomains are expressed as
\begin{eqnarray}\label{eq:MSE-interface}
\mathrm MSE_{U_{\partial \Omega_{i,j}}} = \sum_{\Omega_{i,j}^{+}} \left( \frac{1}{N_{u}} \sum_{k=1}^{k=N_{u}} \left | U^k_{\partial \Omega_{i,j}} - \left(\frac{\widehat{U^k}_{\partial \Omega_{i,j}} + \widehat{U^k}_{\partial \Omega_{i,j}^{+}}} {2}\right) \right | ^{2}\right),
\end{eqnarray}
\begin{eqnarray}\label{eq:MSE-interface-equ}
\mathrm MSE_{\mathrm{Residual}_{\partial\Omega_{i,j}}} = \sum_{\Omega_{i,j}^{+}} \left( \frac{1}{N_{F}} \sum_{k=1}^{k=N_{F}} | R(\widehat {U^k}_{\partial \Omega_{i,j}}) - R(\widehat {U^k}_{\partial \Omega_{i,j}^{+}}) | ^{2} \right)
\mathrm{\,\,and}
\end{eqnarray}
\begin{eqnarray}\label{eq:MSE-flux}
\mathrm MSE_{\mathrm{flux}} = \sum_{\Omega_{i,j}^{+}} \! \left( \frac{1}{N_{F}} \sum_{k=1}^{k=N_{F}} \left | \left(\dfrac{\partial \widehat{U^k}_{\partial \Omega_{i,j}}}{\partial x} + \dfrac{\partial \widehat{U^k}_{\partial \Omega_{i,j}}}{\partial y}\right) \! \cdot \! \hat{\bm{n}} - \!
\left (\dfrac{\partial \widehat{U^k}_{\partial \Omega_{i,j}^{+}}}{\partial x} + \dfrac{\partial \widehat{U^k}_{\partial \Omega_{i,j}^{+}}}{\partial y}\right) \! \cdot \! \hat{\bm{n}} \right| ^{2}\right).
\end{eqnarray}
The term $\mathrm MSE_{U_{\partial \Omega_{i,j}}}$ represents the data-fitting term along the interfaces, $U_{\partial \Omega_{i,j}}$ denotes the simulation data, and $(\frac{\widehat{U^k}{\partial \Omega{i,j}} + \widehat{U^k}{\partial \Omega{i,j}^{+}}} {2})$ is the average of the solutions along the interfaces predicted by two different networks on subdomains $\Omega_{i,j}$ and $\Omega_{i+1,j}$ or $\Omega_{i,j+1}$. Here, $\partial\Omega_{i,j}^{+}$ refers to the interfaces between subdomains. Furthermore, $\mathrm MSE_{\mathrm{Residual}_{\partial\Omega_{i,j}}}$ and $\mathrm MSE_{flux} $ represent the residual continuity conditions and fluxes across common interfaces, respectively. The residuals and fluxes at the interfaces are calculated by two different networks on $\Omega_{i,j}$ and other connected subdomains $\Omega_{i+1,j}$ or $\Omega_{i,j+1}$. For the Allen-Cahn Equation \eqref{eq:modela}, the fluxes in the $x$- and $y$-directions are $\dfrac{\partial \widehat{U}}{\partial x}$ and $\dfrac{\partial \widehat{U}}{\partial y}$, respectively, and $\hat{\bm{n}}$  denotes the direction of the outward normal to the interfaces.  

\begin{figure}[!tbh]
    \centering
    \includegraphics[width= 1\textwidth]{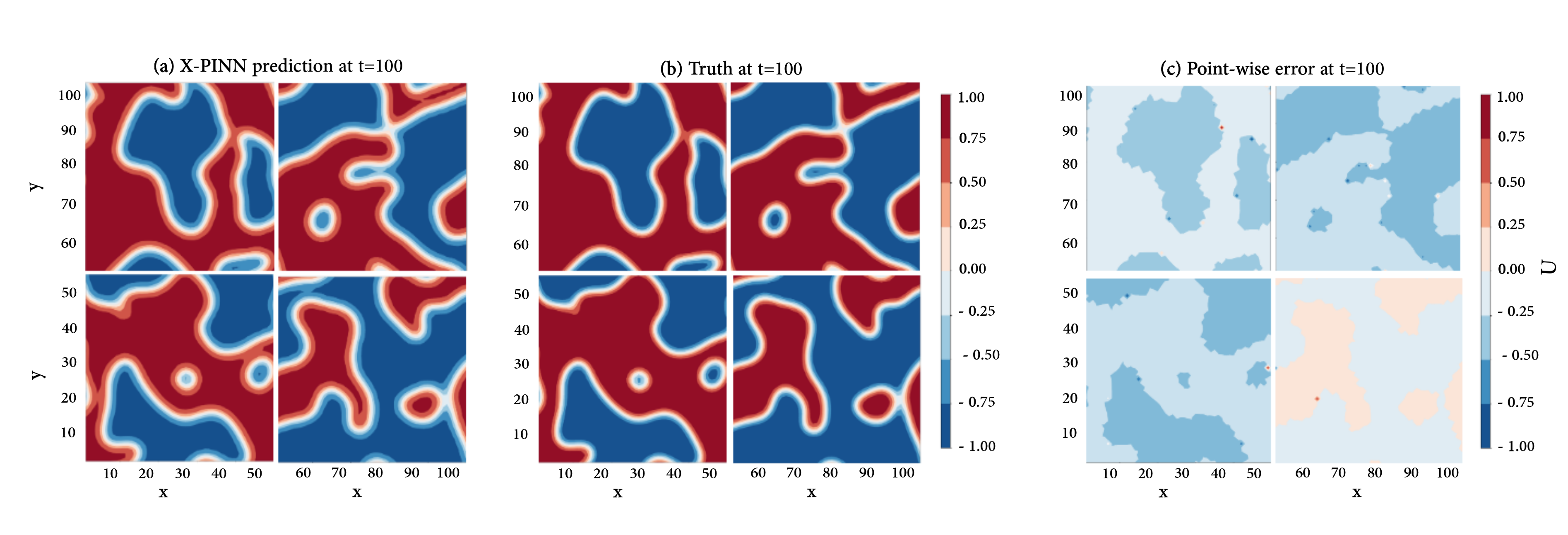}
    \includegraphics[width= 0.49\textwidth]{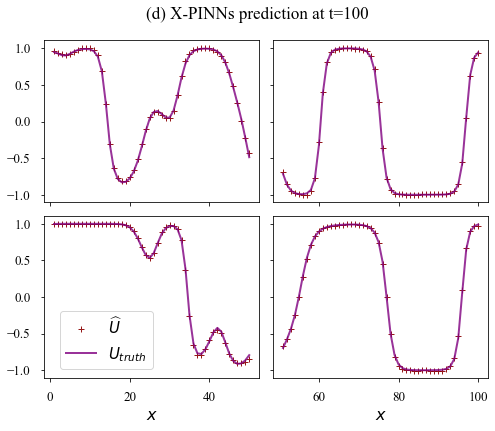}
    \includegraphics[width= 0.49\textwidth]{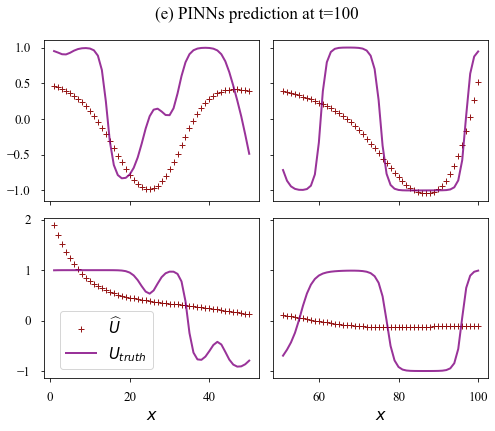}
    \caption{Snapshots of (a) PINNs predictions ${\widehat{U}}$, (b) the true solution of the Allen--Cahn Equation~\eqref{eq:modela} at time $t=100$, and (c) the point-wise relative errors.
    A comparison of the predictions from X-PINNs and PINNs frameworks
    with true values of $U$ is shown in (d) and (e), respectively.}
    \label{fig:Allen-predictions}
\end{figure}

Figure~\ref{fig:Allen-predictions} displays the performance of the trained sub-PINNs. The left panels in Figure~\ref{fig:Allen-predictions}(a) show the contours of the predicted ${\widehat{U}}$ at time $t=100$, while the middle panels in Figure~\ref{fig:Allen-predictions}(b) show snapshots of the exact solution of ${U}$. The right panels in Figure~\ref{fig:Allen-predictions}(c) illustrate the point-wise relative errors. Moreover, Figures~\ref{fig:Allen-predictions}(d) and (e) depict a comparison between the true values with the predicted values of $U$ using X-PINNs and PINNs, the vertical axis represents $U$ and the horizontal axis the $x$-coordinate. The results indicate that the sub-networks in X-PINNs are effective at accurately capturing the primary characteristics of the solution. In contrast, the predictions obtained using PINNs are significantly divergent from the true values, suggesting that PINNs may not be a suitable option for predicting this particular equation. However, some slight errors are present in isolated grid points near sharp boundaries, indicating that the network may not detect smaller features. Such behavior has been previously reported in other studies~\cite{oommen2022learning, kiyani2022machine}.

\begin{figure}[!tbh]
\centering
\includegraphics[width= 1\textwidth]{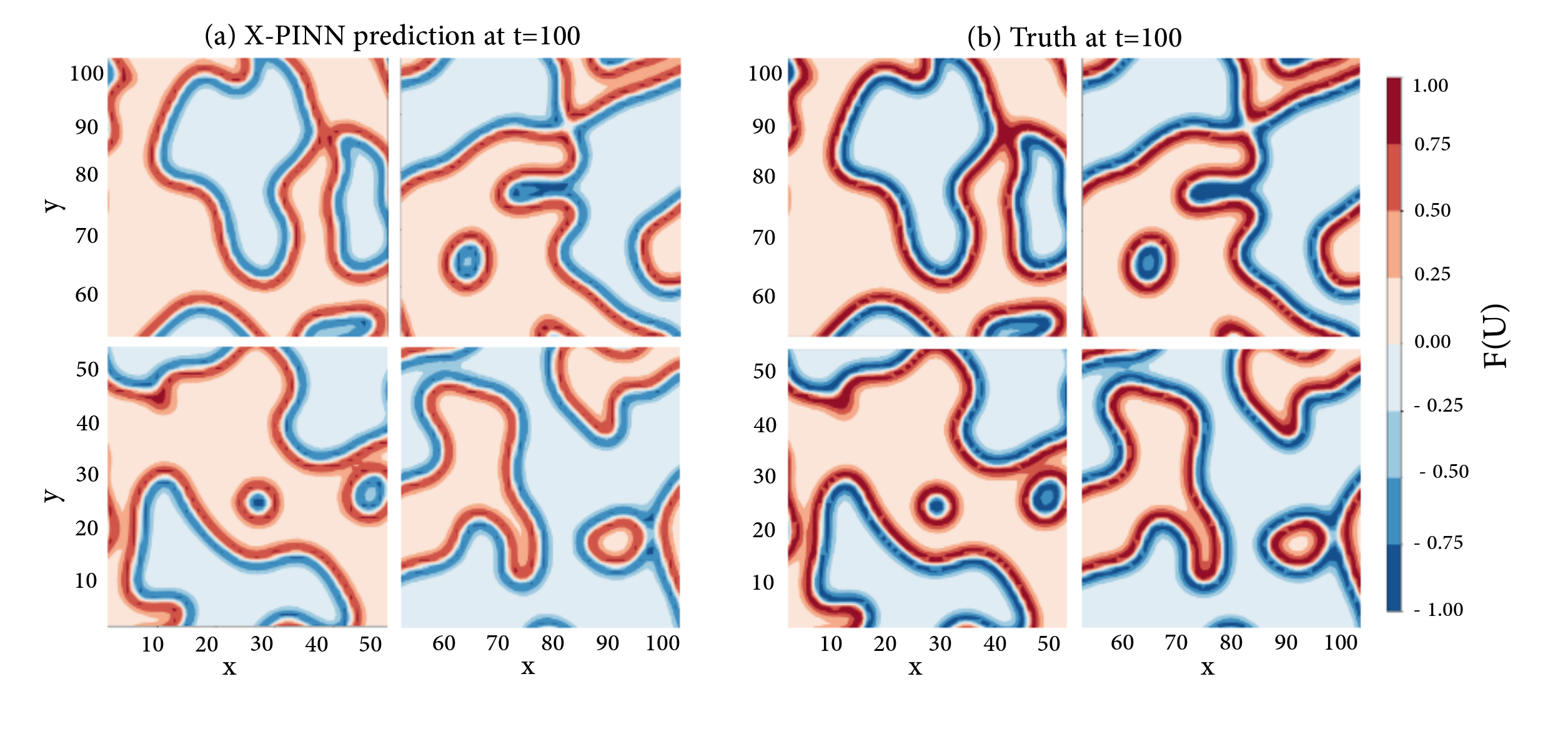}
\includegraphics[width= 0.7\textwidth]{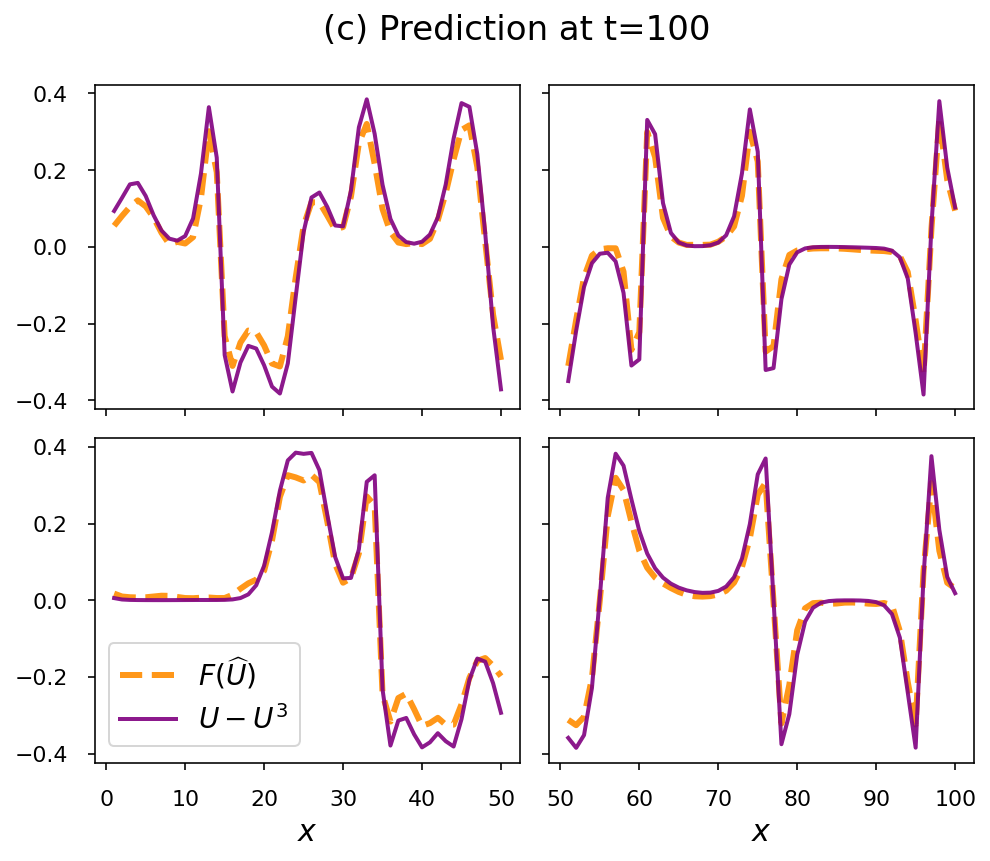}
\caption{Snapshots of (a) predicted unknown function $F(\widehat{U})$, (b) the exact $F(U)=U - U^3$ at time $t=100$
as well as (c) a comparison of the predicted $F(\widehat{U})$ and exact $U - U^3$, where the vertical axis represents $F(\widehat{U})$ and $U - U^3$ and the horizontal axis is the $x$-coordinate.}
\label{fig:F-contour}
\end{figure}

Figure~\ref{fig:F-contour} displays the predicted $F(\widehat{U})$ at time $t=100$. The left panels in Figure~\ref{fig:F-contour}(a) show the contours of $F(\widehat{U})$, while the right panels in Figure~\ref{fig:F-contour}(b) depict $U - U^3$. The results demonstrate that PINNs in each subdomain can accurately identify function $F(U)$ with a high degree of accuracy. Figure~\ref{fig:F-contour}(c) presents a visual comparison between the predicted and exact $F(U) = U - U^3$. The vertical axis represents $F(U)$, and the horizontal axis the $x$-coordinate. These results illustrate that the proposed framework can uncover the unknown part of the nonlinear Equations~\eqref{eq:F} with a great accuracy, despite using no information about the function $F$ during model training.

The primary reason behind the underperformance of PINN compared to X-PINN is spectral bias. Spectral bias in neural network causes it to perform better with good convergence for low frequencies/wavenumbers over high frequencies/wavenumbers. The solution of the Allen-Cahn equation comprises broad range of wavenumbers associated with respective the energy modes of the solutions, and they are difficult to recover for a single PINN due to limited  expressive power of a single neural network. X-PINN, however, comprising multiple PINNs, offers a better representation of the solution by splitting the domain into various subdomains. In this way, each subdomain is represented by a subset of the global wavenumbers and offering rapid to accurate solution.

To evaluate the accuracy of the predictions, we use the Frobenius matrix norm \cite{trefethen2022numerical}  to measure the errors between the predicted $\widehat{U}$ and $F(\widehat{U})$, and the exact values of $U$ and $U- U^3$ in $\mathbb{R}^{100 \times 100}$. The Frobenius matrix norm is commonly used in linear algebra, numerical analysis, and ML. It has several useful properties, including being invariant under orthogonal transformations and sub-multiplicative, similar to the magnitude of a vector. The Frobenius norm for an $m \times n$ predicted matrix $\widehat{U}$ is defined as
\begin{equation}\label{eq:error(U)}
\lVert \text{error}_{U} \rVert_{F} = \sqrt{\sum_{i=1}^{m} \sum_{j=1}^{n} |U_{i,j} - \widehat{U}_{i,j}| ^{2}}.
\end{equation}
Similarly, we calculate the $F(\widehat{U})$ error using 
\begin{equation}\label{eq:error(F)}
\lVert \text{error}_{F(U)} \rVert_{F} = \sqrt{\sum_{i=1}^{m} \sum_{j=1}^{n} |(U_{i,j}-U_{i,j}^3) - F(\widehat{U}_{i,j})| ^{2}}.
\end{equation}

\begin{figure}[!tbh]
\centering
\includegraphics[width=0.494\textwidth]{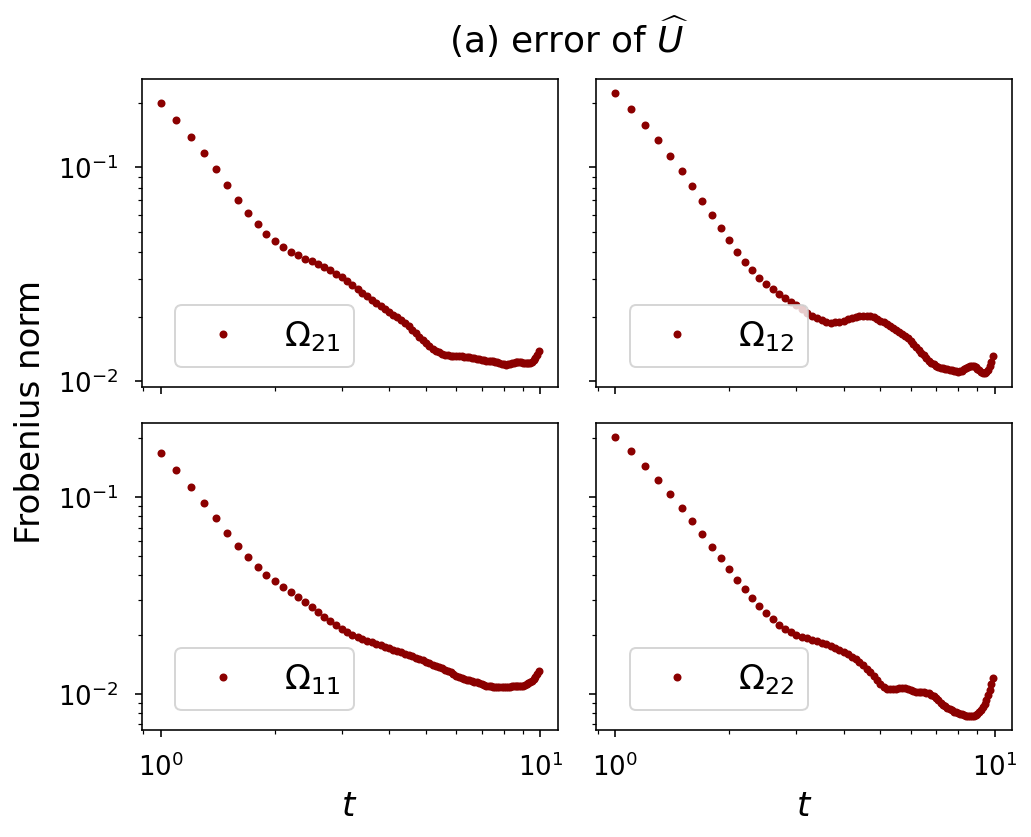}
\includegraphics[width=0.494\textwidth]{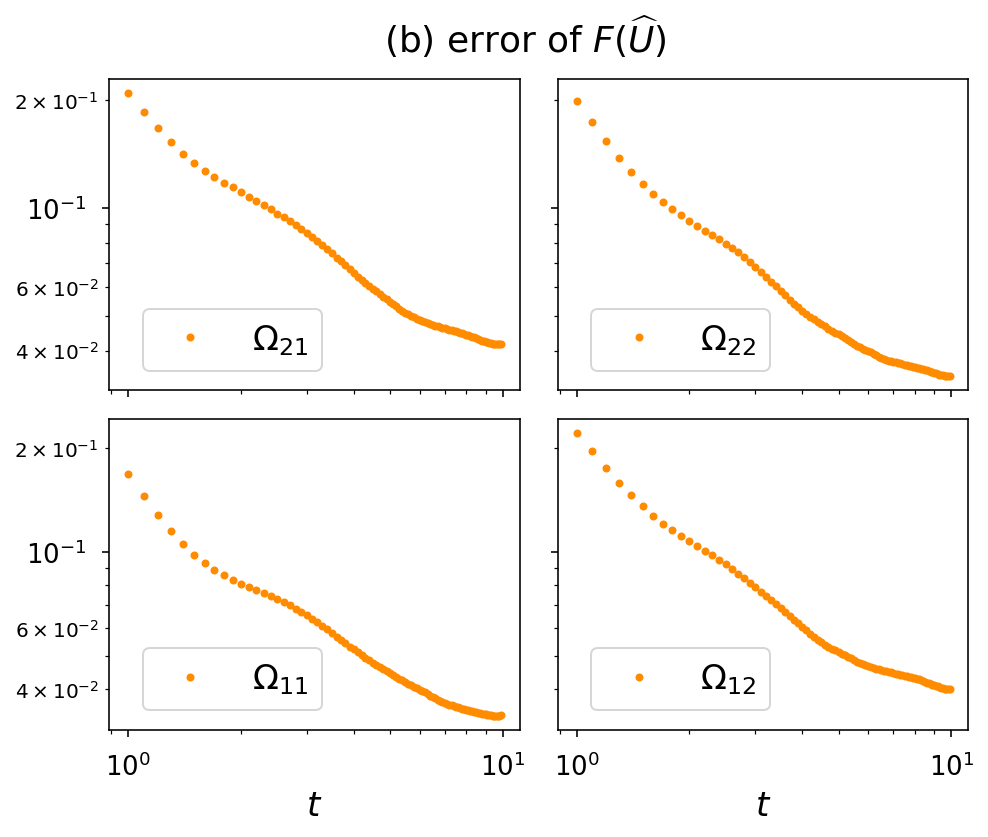}
\caption{Frobenius norm error for (a) $\widehat{U}$ and (b) $F(\widehat{U})$ calculated by Equations \eqref{eq:error(U)} and \eqref{eq:error(F)}.}
\label{fig:error}
\end{figure}

Figure~\ref{fig:error} presents the Frobenius norm errors.  Figure~\ref{fig:error}(a) shows the errors for $\widehat{U}$ and Figure~\ref{fig:error}(b)  for $F(\widehat{U})$. The errors are given as the overall difference between the predicted and exact values of $U$ and $F(U)$ according to Equations~\eqref{eq:error(U)}~and~\eqref{eq:error(F)}. 
The $x$-axis represents the time and the $y$-axis the Frobenius norm error. It can be observed that the error decreases as the model is trained, indicating that the model is learning and makes better predictions as time progresses. This trend is observed in all subdomains. As the model becomes more familiar with the problem, it can identify and capture more complex features of the solution. However, as the error approaches a plateau around $10^{-2}$, the improvement in accuracy slows down. This could be due to several factors such as the model reaching its capacity, the complexity of the problem, or the quality of the training data. Overall, the observed trend in the Frobenius norm error indicates that the model is learning and improving in accuracy.

\begin{figure}[!tbh]
\centering
\includegraphics[width= 0.495\textwidth]{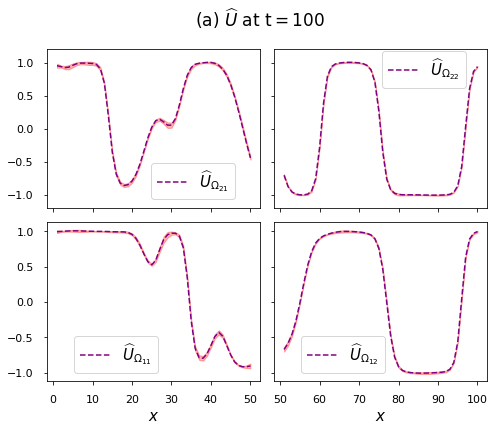}
\includegraphics[width= 0.495\textwidth]{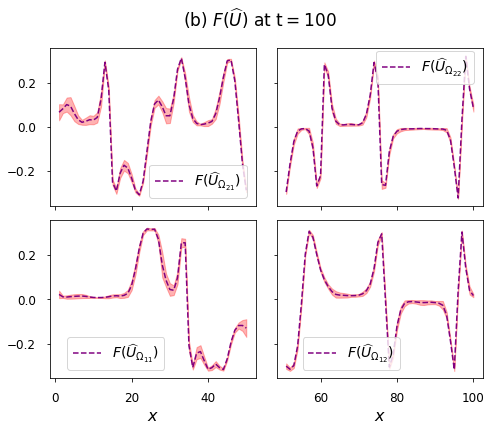}
\caption{The mean value of the predictions (a) $\widehat{U}$ and (b) $F(\widehat{U})$ as well as the standard deviations for the ten different runs. The dashed lines are the averages of the ten different predictions with random selection of training sets. The highlights indicate the standard deviations. Predicted $\widehat{U}$ and $F(\widehat{U})$ are shown in the vertical axis and the horizontal axis is the $x$-coordinate.}
\label{fig:mean-standard}
\end{figure}

In order to investigate the stability and consistency of the trained model, we compute the means and the standard deviations of the predicted solutions across ten runs with different initialization parameters for the neural network. This helps us to determine how much the predicted values vary from their expected values due to the stochastic initialization process. 
Smaller standard deviation indicates that the predicted values are more accurate because their values cluster more tightly around the mean. 
The results are presented in Figure~\ref{fig:mean-standard}, where the left set of figures shows the predicted $\widehat{U}$ and the right one the predicted function $F(\widehat{U})$. The dashed lines represent the mean values, while the highlights above and below the mean indicate the standard deviation along the $x$-axis.

\section{Symbolic regression}\label{Symbolic regression}

Symbolic regression is an ML technique that involves identifying a mathematical expression or an equation that closely approximates a given dataset. This approach has been used in various fields, including engineering, finance, biology, and modeling of complex systems by discovering correlations between input/output data pairs~\cite{billard2002symbolic,vaddireddy2020feature,claveria2017assessment,tuan2006solving,fitzsimmons2018symbolic}. Instead of using predefined models, such as linear or polynomial regression, symbolic regression searches through a space of mathematical functions to find the function that best fits the data. In this study, we utilize the API for symbolic regression provided by the Python library gplearn~\cite{stephens2016genetic} to discover mathematical expressions for $F(\widehat{U})$ as a function of simulation data $U$.

It is worth noticing that here the exact function for $F(U)$ is given by $ U - U^3$, as described by Equations~\eqref{eq:modela} and~\eqref{eq:F}. We set the population size to $5,000$ and evolved $20$ generations until the error became close to $1~\%$. Since the equation $ U - U^3$ consists of basic operations, it does not require the use of custom functions. The results of symbolic regression for multiple runs are presented in Table~\ref{tab:symbolic(U)}, where the predicted function $F(\widehat{U})$ for each subdomain is denoted as $F(\widehat{U}_{\Omega_{i,j}})$ with $i, j \in { 1,2 }$. The findings demonstrate that the model has accurately identified the underlying pattern between the input and output variables, and that the predicted functions contain the correct terms of $U$ and $U^3$ with coefficients that are relatively close to the coefficients (both equal to one) of the equation $ U - U^3$. A comparison between the exact function $F = U - U^3$ and the predicted functions further highlights the effectiveness of X-PINN in discovering the unknown components of equations, and is further validated by symbolic regression. The detailed pseudo-algorithm pairing X-PINN and symbolic regression is shown in Algorithm~\ref{alg:GBAC}.

\begin{table}
\centering
\begin{tabular}{l|l|l|l|l}
\hline
  & $F(\widehat{U}_{\Omega_{11}})$ & $F(\widehat{U}_{\Omega_{12}})$ & $F(\widehat{U}_{\Omega_{21}})$ & $F(\widehat{U}_{\Omega_{22}})$\\ 
\hline
1st run &  $0.96 U (1- 0.89 U^2)$ &  $0.96 U (1- 0.89 U^2)$ & $0.88(1 - U^{3}) $ & $0.96 U (1- 0.89 U^2)$ \\ 

2nd run & $0.88U(0.99 - U^2)$& $0.96 U (1- 0.89 U^2)$&  
$0.88U(0.994 - U^2)$&  $0.96 U (1- 0.89 U^2)$\\

3rd run & $0.96 U (1- 0.89 U^2)$&  $0.96 U (1- 0.89 U^2)$& $0.88U(0.99 - U^2)  $ &   $ 0.88 U(1 -  U^2)$ \\

4th run & $0.88 U (0.99 - U^2)$ & $0.88 U (0.99 - U^2)$& $ 0.96 U (1- 0.89 U^2)$ & $0.88 U(0.99 - U^2)$  \\

5th run & $0.88 U(0.994 - U^2)$ & $0.88 U (0.994 - U^2)$& $ 0.88 U(0.99 - U^2)$ & 
$ 0.96 U (1- 0.89 U^2)$  \\
\hline
\end{tabular}
\caption{Symbolic regression results for multiple runs to fit a mathematical formulation to the predicted function $F(\widehat{U}_{\Omega_{i,j}})$ on subdomain $_{\Omega_{i,j}}$. It is worth noting that the exact formulation for $F(\widehat{U})$ is $U-U^3$, and this formulation was closely approximated in multiple runs.}
\label{tab:symbolic(U)}
\end{table}
%


\section{Noisy data analysis}\label{Noisy data}

To demonstrate the robustness of the proposed framework to discover equations from noisy data, we 
add noise to the original datasets to evaluate the performance of the framework. The purpose of adding noise to the data is to simulate real-world scenarios where data is often subject to randomness (e.g. thermal noise) and errors, rather than being precise and clean. Uniform noise of various magnitudes, sampled from a uniform distribution of zero mean and unit standard deviation, was applied to the original data set, which was used for training the PINNs.

\begin{figure}
\centering
\includegraphics[width=0.45\textwidth]{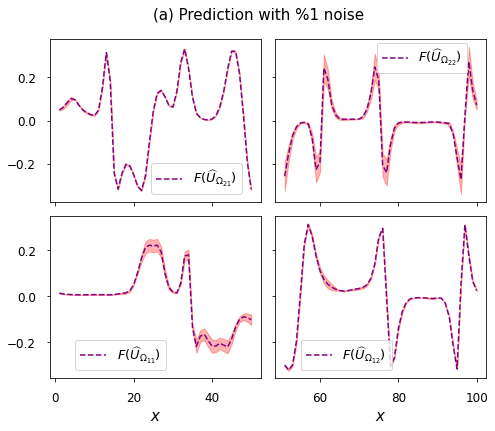}
\includegraphics[width=0.45\textwidth]{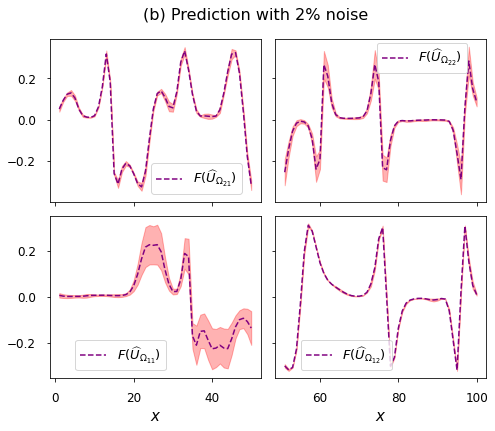}
\includegraphics[width=0.45\textwidth]{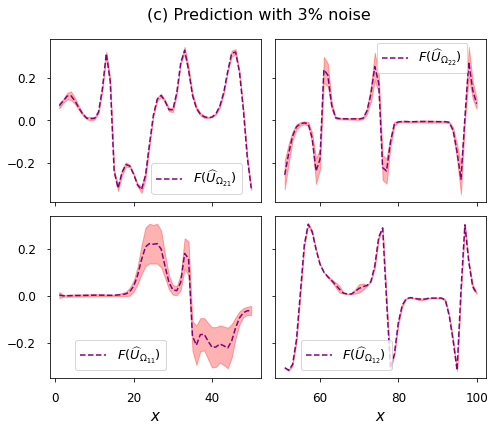}
\includegraphics[width=0.45\textwidth]{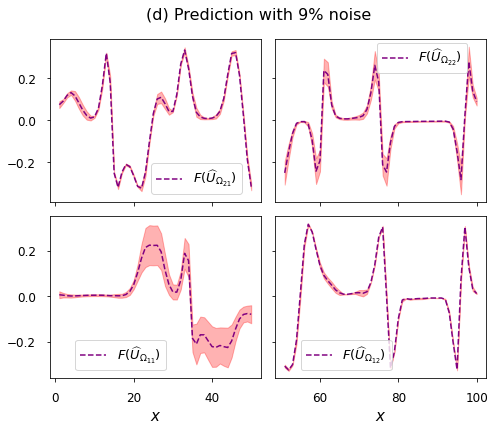}
\caption{Mean and standard deviation of the predicted $F(\widehat{U}_{\Omega_{ij}})$ for each subdomain $\Omega_{ij}$  derived from noisy data with (a) $1\%$ noise, (b), (c), and (d) $2 \%$, $3 \%$, $9 \%$ noise respectively. The vertical axis represents $F(\widehat{U}_{\Omega_{ij}})$ values, while the horizontal axis represents the corresponding values of $x$. The dashed lines represent the mean values of the predicted $F(\widehat{U}_{\Omega_{ij}})$, while the red highlights around the dashed lines indicate the standard deviation.}
\label{fig:noise}
\end{figure}

A summary of results using noisy data is presented in Figure~\ref{fig:noise}. The predicted $F(\widehat{U})$ is shown in Figure~\ref{fig:noise}(a) with $1\%$ noise, while Figures~\ref{fig:noise}(b), (c), and (d) present the results for $2\%$, $3\%$, and $9\%$ noise, respectively. 
The plots show the mean and standard deviation values that were calculated across ten runs for each set of predictions.
Our results demonstrate that the sub-PINNs were able to accurately identify the unknown function $F$ from noisy data. Notably, predictions on subdomains $\Omega_{12}$ and $\Omega_{21}$ were found to be less sensitive to noise than the other two subdomains. Additionally, we observe that the subdomain $\Omega_{11}$ has a higher standard deviation, indicating greater sensitivity to noise.

As above, we employ symbolic regression to obtain the mathematical formula for the predicted function $F(\widehat{U})$ using noisy data, and compare it with the exact expression $U-U^3$. The predicted $F(\widehat{U})$ and $U$ were fed into the symbolic regression model and the outcomes are presented in Table~\ref{tab:noise}. The results show that the algorithm has successfully identified the correct terms of $U$ and $U^3$.

\begin{table}
\centering
\begin{tabular}{l|l|l|l|l}
\hline
  & $F(\widehat{U}_{\Omega_{11}})$ & $F(\widehat{U}_{\Omega_{12}})$ & $F(\widehat{U}_{\Omega_{21}})$ & $F(\widehat{U}_{\Omega_{22}})$\\ 
\hline
($1 \%$ noise)  & $0.96 U (1- 0.89 U^2)$ &  $ 0.96 U (1- 0.89 U^2) $ & $ 0.87 U (1 - U^2)$ & $0.96 U (1- 0.89 U^2)$ \\ 
\hline
($2 \%$ noise)  & $0.96 U (1- 0.89 U^2)$ &  $  0.96 U (1- 0.89 U^2) $ & $U (0.96  - 0.98 U^2) $ & $0.96 U (1- 0.89 U^2)$ \\ 

\hline
($3 \%$ noise) & $ 0.96 U (1- 0.89 U^2)$ & $0.96 U (1- 0.89 U^2)$ &  $ 0.956 U (1.02 - U ^ 2)$ & $ 0.956 U (1- 0.85 U^2)$  \\ 
\hline

\hline
($9 \%$ noise) & $ 0.96 U (1- 0.89 U^2) U$ & $0.956 U (1- 0.89 U^2)$ &  $ 0.96 U (1.02 - U ^ 2)$ & $ 0.96 U (1- 0.89 U^2)$  \\ 
\hline
\end{tabular}
\label{tab:noise}
\caption{Symbolic regression results for noisy data. $F(\widehat{U}_{\Omega_{i,j}})$ is the predicted function corresponding to subdomain $\Omega_{i,j}$. 
It is noteworthy that the exact formulation for $F(\widehat{U})$ is given by $U-U^3$. Despite the presence of $9 \%$ noise in the dataset, the results demonstrate that the algorithm has successfully identified a closely approximated formulation for the predicted $F(\widehat{U})$.}
\end{table}

\begin{figure}
\centering
\includegraphics[width=0.495\textwidth]{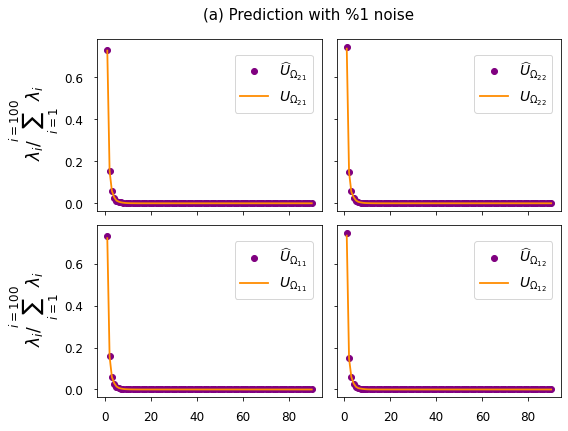}
\includegraphics[width=0.495\textwidth]{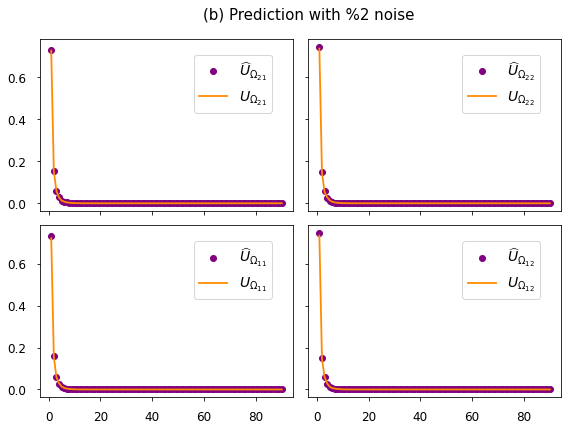}
\includegraphics[width=0.495\textwidth]{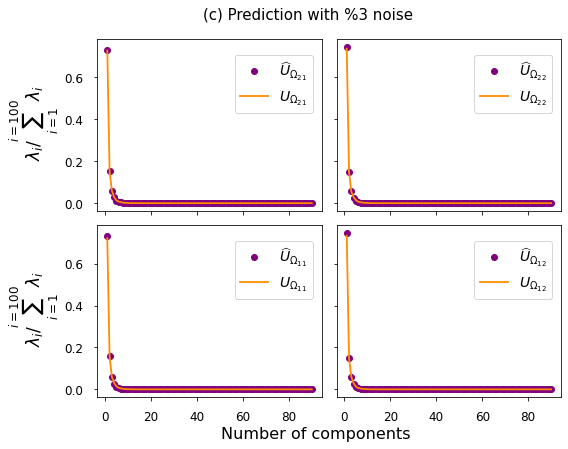}
\includegraphics[width=0.495\textwidth]{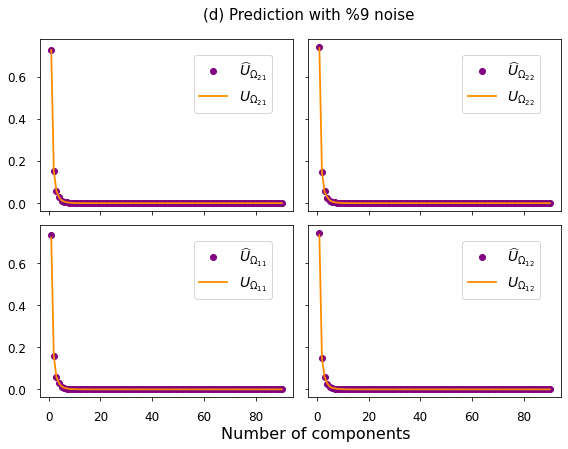}
\caption{The ratio of each singular value to the sum of all singular values for the predicted $\widehat{U}_{\Omega_{ij}}$ for each subdomains $\Omega_{ij}$ with (a) $1\%$ noise, (b), (c), and (d) $2 \%$, $3 \%$, $9 \%$ noise respectively. The vertical axis represents $ \lambda_{i} / \sum_{i=1}^{i=100}\lambda_{i}$ values, while the horizontal axis represents the number of samples. The ratio of singular values decreases, which suggests that the informative singular vectors are becoming less dominant or relevant in predicting the target variable.}
\label{fig:noise-eig}
\end{figure}

To evaluate the model's performance in recovering all the energy modes of system, we performed a modal analysis using proper orthogonal decomposition (POD)~\cite{Weiss2019-wt}.
Figure~\ref{fig:noise-eig} shows the ratios of the singular values of the matrices $U$ (modal energy) and $\widehat{U}$ that have been reshaped to ($n_{x} \times n_{y} , n_{t}$) to the sum of all singular values (total energy of the system). The singular values were obtained by performing  singular value decomposition (SVD) on the matrices $U$ and $\widehat{U}$. Specifically, the SVD of the matrix $U$ is computed using the equation $U = W \Sigma V^{T}$, where $W$ and $V$ are orthogonal matrices, and $\Sigma$ is a diagonal matrix containing the singular values of $U$ on the diagonal~\cite{ziegel2003elements, wall2003singular}. 
Noise that is uncorrelated and evenly distributed across the entire dataset can lead to instability in predictions or model performance. Therefore, the ratio of each singular value to the sum of all singular values can be used to assess the stability of predictions. As shown in Figure~\ref{fig:noise-eig}, the ratios of singular values decrease over samples, indicating that the informative singular vectors are becoming progressively less dominant, or relevant, in the prediction task. Figures~\ref{fig:noise-eig}(a)-(d) show that the model is able to recover the dominant energy modes very accurately even when the level of the noise is increased. 
%
%


\section{Optimal Training Datasets}\label{Reducing training data sets}

In this section, we reduce the number of training points while still preserving its overall shape and distribution to understand how much information is needed to train a network. The technique is called downsampling, which involves reducing the number of training points without sacrificing too much accuracy or losing too much information.
The goal is to enhance the generalization performance of the model on unseen data and to reduce its complexity by minimizing the number of training points without compromising accuracy or losing crucial information.
For this purpose, we choose a downsampling rate that is appropriate for the dataset and ensure that the selected subset of data is representative of the entire dataset. To select an appropriate downsampling rate, one can consider factors such as the size of the original dataset, the complexity of the problem, and the computational resources available. It is also important to ensure that the selected subset of data is representative of the entire dataset, meaning that it captures the overall distribution and patterns of the original data. To evaluate the performance of the networks, we randomly select samples from the original dataset and consider four different sizes to ensure that the downsampling rate is not too high, resulting in a loss of critical information or a decrease in accuracy.

In the previous sections, our datasets were randomly split into training and test with a $80 : 20$ ratio. Each subdomain has thus been trained by using $80$ samples, which implies a training set of size $(80, 50, 50)$, where $80$ is the number of samples, and each subdomain has dimensions of $n_{x}=n_{y}=50$.

\begin{figure}[!tbh]
\includegraphics[width=1\textwidth]{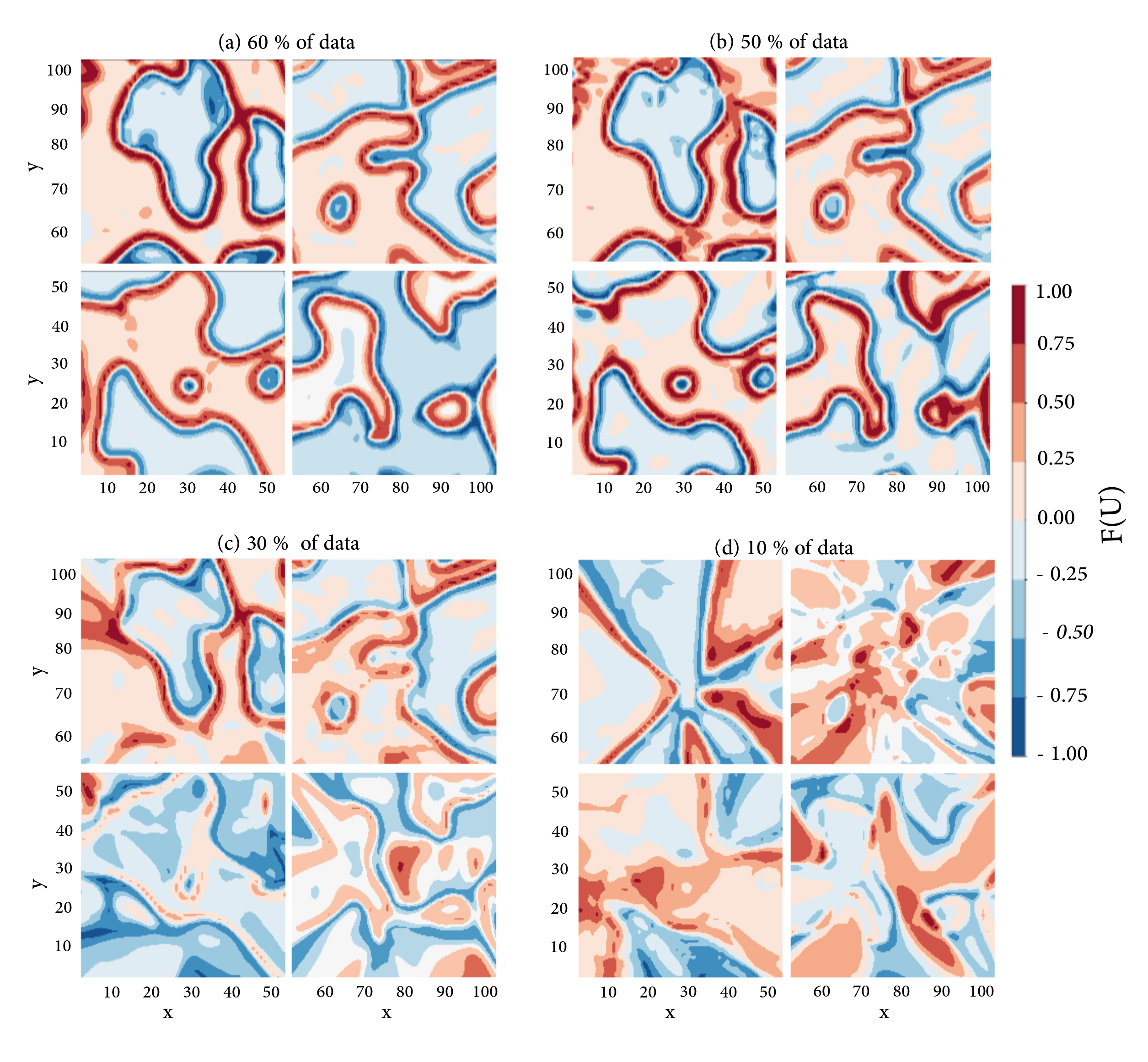}
\caption{Snapshots of the predicted $F(\widehat{U})$ obtained from varying sample sizes of data. (a) illustrates the predicted $F(\widehat{U})$ when $60\%$ of the data was used for training. (b), (c), and (d) correspond to cases where $50 \%$, $30\%$, and $10\%$ of the data were used for training, respectively. The results indicate that the neural networks can effectively capture the underlying information in (a) and (b) using only $60 \%$ and $50 \%$ of the original training data, respectively. However, in the case of (c) and (d), which have a much smaller subset of training points, the network is not able to perform well due to the lack of sufficient training data.}
\label{fig:down-sampling1}
\end{figure} 

\begin{figure}
\centering
\includegraphics[width=0.495\textwidth]{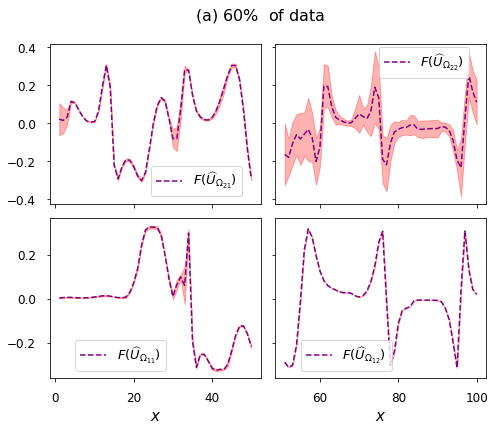}
\includegraphics[width=0.495\textwidth]{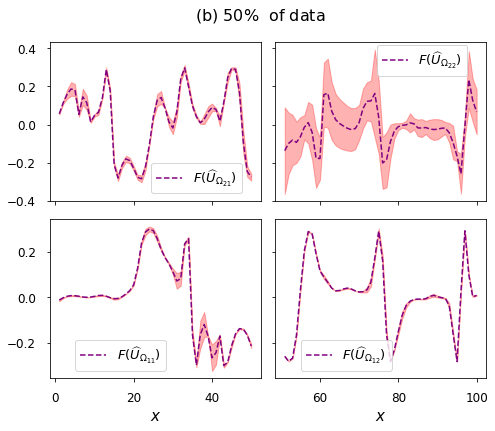}
\includegraphics[width=0.495\textwidth]{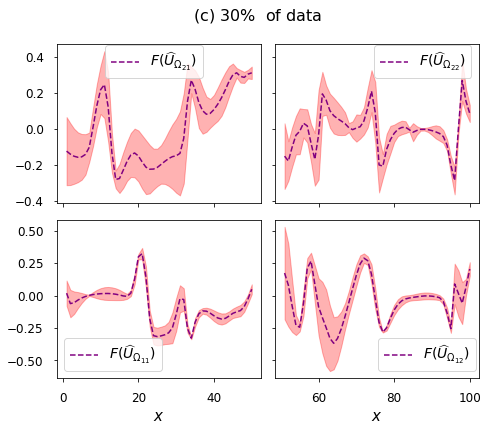}
\includegraphics[width=0.495\textwidth]{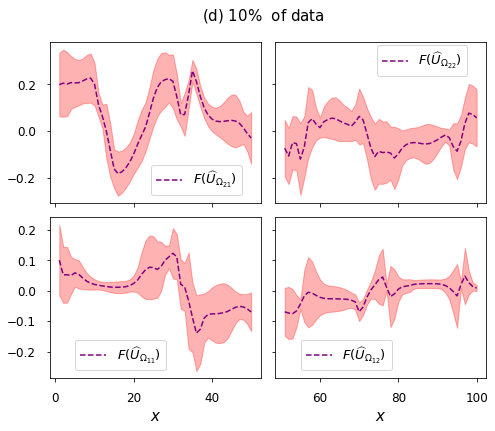}
\caption{The standard deviation of predicted $F(\widehat{U}_{\Omega_{ij}})$ for each subdomains $\Omega_{ij}$ was computed using various sets of training points across multiple runs. The vertical axis represents $F(\widehat{U}_{\Omega_{ij}})$, while the horizontal axis represents $x$. In (a), the predicted $F(\widehat{U})$ was generated using $60\%$ of the available training points, while (b), (c), and (d) show the predicted $F(\widehat{U})$ for models trained with only $50 \%$, $30 \%$, $10 \%$ of the training points, respectively.
The plots depict the mean values of the predicted $F(\widehat{U}_{\Omega_{ij}})$ using dashed lines. The red highlights surrounding the dashed lines indicate the standard deviation of ten runs.}
\label{fig:down-sampling2}
\end{figure}

To reduce the number of samples, we decreased the size of training data by reducing the number of training points in both time and space as illustrated in Figures~\ref{fig:down-sampling1} and~\ref{fig:down-sampling2}.
Figure~\ref{fig:down-sampling1}(a) indicates that for each subdomain, a training set of size $(60,30,30)$ was selected to represent $60\%$ of the available data. As depicted in Figure~\ref{fig:down-sampling1}(b), (c), and (d), we trained the model using $50\%$, $30\%$, and $10\%$ of the data, respectively. This corresponds to training sets of sizes $(50,25,25)$, $(30,15,15)$, and $(10,5,5)$ for each subdomain. The findings demonstrate that the networks can effectively capture the information in Figures~\ref{fig:down-sampling1}(a) and (b).

In contrast, in Figures~\ref{fig:down-sampling1}(c) and (d) the model failed to accurately identify the unknown function, suggesting a lack of sufficient training. Hence, at minimum, $50\%$ of the data must be used to train the network to adequately capture the information.

We present the standard deviations and averages of the predicted $F(\widehat{U})$ for ten runs in Figure~\ref{fig:down-sampling2} to evaluate the sensitivity of the model's predictions to changes in the training data. Figure~\ref{fig:down-sampling2}(a) shows the predicted $F(\widehat{U})$ computed using $60\%$ of the available training data, and Figures~\ref{fig:down-sampling2}(b), (c), and (d) show the predicted $F(\widehat{U})$ for models trained with $50\%$, $30\%$, and $10\%$ of the training data, respectively.

The results in Figures~\ref{fig:down-sampling2}(a) and (b) indicate that the subdomains $\Omega_{11}$, $\Omega_{12}$, and $\Omega_{21}$ exhibit low sensitivity (i.e., low standard deviation), that is, they are less affected by changes in the training data and are more likely to generalize well to new data. However, in $\Omega_{21}$, higher sensitivity was observed, which could indicate reduced robustness and generalization ability to new, unseen data. In contrast, in Figures~\ref{fig:down-sampling2}(c) and (d), all subdomains exhibited high sensitivity, with a high standard deviation around mean predictions. This suggests that the model is highly sensitive to changes in the training data, which could affect its ability to generalize to new and unseen data.
\begin{table}
\centering
\begin{tabular}{l|l|l|l|l}
\hline
  & $F(\widehat{U}_{\Omega_{11}})$ & $F(\widehat{U}_{\Omega_{12}})$ & $F(\widehat{U}_{\Omega_{21}})$ & $F(\widehat{U}_{\Omega_{22}})$\\ 
\hline
($60 \%$ data)  & $0.96 U (1- 0.89 U^2)$ &  $ 0.96 U (1- 0.89 U^2)$ & $ 0.96 U (1- 0.89 U^2)$ & $0.96 U (1- 0.89 U^2)$ \\ 
\hline
($50 \%$ data)  & $0.96 U (1- 0.89 U^2)$ &  $ 0.56 U (1- 0.144 U^2) $ &
$ 0.56 U (1- 0.144 U^2)$ & $ 0.96 U (1- 0.89 U^2)$ \\ 

\hline
($30 \%$ data)  & $0.088 U$ &  $0.088 U$ & $0.088 U$ & $0.088 U$ \\ 
\hline
($10 \%$ data)  & $ 0.04 U $ &  $ 0.02 U $ & $ 0.04 U$ & $ 0.02 U$ \\ 
\hline
\end{tabular}
\label{tab:down-sampling}
\caption{ The mathematical formula for $F(\widehat{U}_{\Omega{ij}})$ for each subdomain $\Omega_{ij}$ using symbolic regression. The model was trained using different percentages of the available data, and the results show that the correct terms of $U$ and $U^3$ were accurately identified when the model was trained with (a) $60 \%$ and (b) $50 \%$ of the data. However, the model's accuracy decreased with less data, and it could not accurately predict coefficients when trained with only $50 \%$ of the data. Moreover, the results demonstrate that training the model with only $30 \%$ and $10 \%$ of the data is insufficient to even predict the correct terms of $U$ and $U^3$ in the function. }
\end{table} 

The mathematical expression of the function $F(\widehat{U})$ was discovered using a symbolic regression model by feeding in the predicted $F(\widehat{U})$ for different training data set sizes. The outcomes of the symbolic regression model are summarized in Table~\ref{tab:down-sampling}. The model was able to identify the correct terms of $U$ and $U^3$ for the cases where $60 \% $ and $50 \% $ of the data were used for training. However, it was found that training the model with only $50 \% $ of the data is not sufficient for predicting the coefficients accurately. In contrast, the results for the cases where only $30 \% $ and $10 \% $ of the data were used for training indicate that the model could not even predict the correct terms of $U$ and $U^3$ in the function. Therefore, based on these results, it is recommended to use at least $60 \% $ of the available datasets for training in order to obtain the correct form of the equation and good approximations for the coefficients in the symbolic regression model.

\section{Summary}\label{Conclusion}

In this study, we have applied modified X-PINNs to perform gray-box learning of the Allen-Cahn equation by decomposing the computational domain into four subdomains. We assumed that the equation comprises a Laplacian component, and our objective was to discover the nonlinear component of the equation, that is, $U - U^3$. The results indicate that X-PINNs have highly expressive capabilities of uncovering the unknown components across various subdomains. Given the X-PINNs' capability to handle complex data, they have the potential to become a valuable tool for gray-box learning of equations. 

In addition, we implemented symbolic regression with the aim to uncover the general mathematical (closed) form of the unknown component (here, $U - U^3$) and its coefficients. The results show great agreement with the exact form of the equation. This outcome serves as a strong evidence of the efficacy of our proposed framework in accurately identifying unknown components of an equation from data/noisy data.

Additionally, we determined the number of data samples required for  training a neural network to accurately identify the unknown component of the equation. Our results show that at least $60 \%$ of the available data is necessary to accomplish that.
 Training with a smaller percentage of data resulted in less accurate and less close-to-exact predicted results. We also demonstrated the robustness of the trained model by computing the epistemic uncertainty using different initializations of the neural network parameters. 
 
 Overall, the findings of this study show that the proposed framework is a powerful tool for discovering the unknown components of nonlinear and complex PDEs by using the domain decomposition approach. Future research could explore the potential of the proposed framework to investigate how to further optimize the performance in different subdomains and using the method for solving PDEs involving higher order derivatives e.g., bi-Laplacians. 

\section*{Acknowledgments}

EK thanks Dr. Aniruddha Bora, Dr. Ehsan Kharazmi, and Zhen Zhang for their invaluable suggestions throughout the different phases of the project, and Mitacs Globalink Research Award Abroad and Western University's Science International Engagement Fund Award. MK thanks the Natural Sciences and Engineering Research Council of Canada (NSERC) and the Canada Research Chairs Program. The work was partially supported by DOE grant (DE-SC0023389). Computing facilities were provided by the Digital Research Alliance of Canada (https://alliancecan.ca). This research was partially conducted by using computational resources and services at the Center for Computation and Visualization, Brown University. 

%



%

\end{document}